*Article*

# Location-Robust Cost-Preserving Blended Pricing in Multi-Campus AI Data Centers

**Qi He**

**Abstract**

Large-scale AI data center portfolios procure identical hardware and service SKUs across geographically diverse campuses, yet finance and operations often require a single system-level "world price" per SKU for budgeting, chargeback, and benchmarking. Deployment-weighted blending preserves total cost but can produce Simpson-type aggregation bias, reversing global SKU rankings due to SKU-specific location mixes and weakening decision signals. This study formalizes the objective of location-robust, cost-preserving blended pricing and develops two implementable operators for production cost pipelines. The first applies a two-way fixed-effects decomposition to separate global SKU effects from campus premia, followed by scalar normalization for exact cost preservation and improved interpretability. The second solves a constrained common-weight optimization to find a single set of nonnegative campus weights that preserves total cost while reducing ranking instability, with feasibility diagnostics and a slack-based fallback when constraints are tight. Simulations and an AI data center OPEX illustration show that the proposed operators substantially reduce ranking violations relative to naïve blending while preserving portfolio cost to numerical precision. The resulting detect-correct-validate workflow provides a scalable decision-support component for cost governance and robust technoeconomic benchmarking in multi-campus cloud infrastructure.

## 1. Introduction

Large-scale cloud and AI providers increasingly operate multi-campus data center portfolios, where identical hardware and service SKUs are procured, deployed, and accounted for across geographically heterogeneous sites. In this environment, finance, procurement, and capacity-planning teams require a single system-level benchmark ("world price") per SKU to support data-driven decision-making in chargeback, budgeting, benchmarking, procurement negotiations, and workload reallocation planning. Constructing such representative prices resembles classic problems in index-number construction and aggregation, where a single measure must summarize heterogeneous local price-quantity pairs while remaining meaningful for operational decisions. The index-number literature emphasizes that aggregation is not purely mechanical: different operators can satisfy accounting identities yet yield materially different implied relative prices and, consequently, different managerial conclusions. [1]-[4]

A naive and widely used baseline is deployment-weighted blending of campus-level unit prices. Although this baseline preserves total cost by construction, it is vulnerable to location heterogeneity: relative SKU rankings can reverse under aggregation even when each campus exhibits consistent within-campus ordering. This Simpson-type phenomenon arises from non-comparable, SKU-specific location mixes, cross-campus shifts in deployment weights contaminate comparisons. [5]-[7] In multi-campus AI infrastructure, such reversals are not statistical curiosities; they can distort decision signals by misguiding procurement trade-offs, mis-prioritizing capacity upgrades, biasing "what-if" workload rebalancing analyses, and undermining fairness in chargeback systems. At the same time, any benchmark used in finance-facing workflows must remain strictly cost-preserving to reconcile with accounting systems and OPEX reporting, especially as power and cooling become first-order drivers of AI data center costs [8]-[13], with emerging integrated solutions exploring waste-to-energy synergies [18].

This paper contributes to decision-making and decision support systems (DSSs) in operational research and management science by formalizing location-robust, cost-preserving blended pricing for multi-campus AI data centers, consistent with recent work on integrated socio-economic metric frameworks in applied computation [21]. The design objective is to construct a world-price vector that (i) exactly preserves system-level cost, (ii) delivers stable and interpretable SKU rankings under location heterogeneity, and (iii) supports scalable implementation in production data pipelines. Table 1 positions common approaches against these requirements. Building on index-number principles [1]-[4] and the mechanism underlying Simpson-type reversals [5]-[7], I propose two complementary decision-support operators that bridge accounting correctness and ranking robustness:

1. A two-way fixed-effects (FE) operator, which decomposes observed prices into global SKU effects and campus premia, followed by scalar normalization for exact cost preservation. This yields interpretable and explainable benchmarks, with natural accommodation of mild missingness via model-based smoothing.
2. A constrained convex common-weight operator, which derives a unified set of nonnegative campus weights via quadratic programming [14], enforcing a conservative location-robust benchmark that eliminates dominance reversals by construction, with built-in feasibility diagnostics and a slack-based fallback.

To operationalize these operators as DSS components, I introduce diagnostic metrics, order violation rate (OVR) and cost distortion ratio (CDR), that quantify reversal risk and enable a structured "detect → correct → validate" workflow. The methods are evaluated through campus-level simulations and a realistic AI data center OPEX case study, demonstrating substantial reductions in ranking violations relative to naive blending while maintaining cost accuracy to machine precision. The proposed system integrates seamlessly into distributed data processing frameworks commonly used in large-scale cloud computing environments. [15] This framing aligns with foundational concepts in decision support systems, which integrate data, models, and user interfaces to assist complex managerial decisions [20].

**Table 1.** Comparison of common price-index / aggregation approaches and the proposed operators

| Approach | Objective | Data | Weights | P1 Cost-preserving | P2 Dominance-robust | P3 Distributed | Main issue |
|---|---|---|---|---|---|---|---|
| Classical / superlative indices [1], [2] | COLI/inflation | 1 vector per period/region | product-specific | not targeted | not targeted | easy | mix-driven rank flips |
| CPI/PPP manuals [3] | comparability | baskets/regions | basket/region | not required | not SKU-focused | moderate | objective mismatch |

| FE / hedonic regressions | adjust effects | micro + covariates | model-implied | not automatic | no guarantee | moderate | interaction risk |
|---|---|---|---|---|---|---|---|
| Naive deployment-weighted blend | chargeback shortcut | price matrix + quantities | SKU-specific mix | yes (exact) | no [5]-[7] | yes | Simpson reversals |
| Method 1: two-way FE world prices | net out campus level | price matrix + quantities | FE + shift | yes | improves (fit-dependent) | yes | fails if strong interactions |
| Method 2: convex common weights | common weights + identity | price matrix + quantities | common simplex | yes | yes in dominance | yes | needs small QP |

## 2. Problem Formulation and Simpson's Paradox

*2.1 Setup and notation*

I consider a single cross-sectional snapshot of a multi-location production system. Products (or planning items) are indexed by $i = 1, \dots, I$, and locations (campuses) by $j = 1, \dots, J$. In data center applications, a 'product' may correspond to a standardized server tray, machine type, or other capacity SKU deployed across multiple campuses. For each pair $(i, j)$ I observe:

- a location-specific unit price $p_{ij} \in \mathbb{R}_+$, and
- a deployed quantity $q_{ij} \in \mathbb{R}_+$.

The total quantity of product $i$ is $Q_i \equiv \sum_{j=1}^{J} q_{ij}$, and the realized total cost in the system is

$$C \equiv \Sigma_i \Sigma_j \, p_{ij} q_{ij}.$$

The firm wishes to replace the full price matrix $\{p_{ij}\}$by a blended world price $\bar{p}_i$for each product $i$. Intuitively, $\bar{p}_i$should summarize the location-specific prices of product $i$into a single representative cost measure that can be used for budgeting, pricing, and high-level scenario analysis.

Formally, I view any mapping

$$\mathcal{A}: \{(p_{ij}, q_{ij})\}_{i,j} \mapsto \{\bar{p}_i\}_{i=1}^{I}$$

as a blended pricing operator. Different choices of $\mathcal{A}$correspond to different aggregation rules.

A natural benchmark is the naive quantity-weighted average, which uses each product's own deployment mix as weights:

$$\bar{p}_i^{(naive)} \equiv \sum_{j=1}^{J} \pi_{ij} \, p_{ij}, \pi_{ij} \equiv \frac{q_{ij}}{Q_i}.$$

This operator is simple, intuitive, and widely used in practice. However, as I discuss below, it can lead to severe ranking reversals of the Simpson type when products have different location mixes.

*2.2 Desired properties of blended pricing operators*

I focus on three core properties that a blended pricing operator $\mathcal{A}$should satisfy in the multi-location cost setting.

- P1. Cost Preservation.

The blended prices $\{\bar{p}_i\}$should preserve total cost when applied to observed quantities:

$$\Sigma_i \bar{p}_i Q_i \ = \ C \ = \Sigma_i \Sigma_j \, p_{ij} q_{ij}.$$

This requirement ensures that replacing the location-specific prices $\{p_{ij}\}$by blended prices $\{\bar{p}_i\}$does not distort total cost at the system level. For internal accounting, profitability analysis, and unit-cost calculations, exact cost consistency is highly desirable.

- P2. Location Robustness in Pairwise Comparisons.

Consider two products $i$and $k$. Suppose that for every campus $j$,

$$p_{ij} \leq p_{kj},$$

with strict inequality for at least one $j$. Locally, product $i$ is weakly cheaper than product $k$at every location. A reasonable aggregation rule should avoid systematic reversals of this ordering at the blended level, a concern that is well known in settings with highly skewed objectives [17]. I therefore seek operators $\mathcal{A}$ that minimize or eliminate cases in which

$$\bar{p}_i > \bar{p}_k$$

despite $p_{ij} \leq p_{kj}$for all $j$. This is a location-robustness requirement: blended prices should, as far as possible, respect consistent local dominance relationships and not manufacture Simpson-type paradoxes.

While P2 cannot be satisfied globally for all possible data matrices $\{p_{ij}, q_{ij}\}$by any non-trivial operator, it provides a benchmark for comparing aggregation rules: operators that generate fewer such reversals under realistic deployment patterns are preferred.

- P3. Implementability in Distributed Systems.

A practical operator should be computable from low-dimensional aggregates and small auxiliary steps, rather than requiring repeated full-matrix operations. Concretely, it should be implementable using distributed group-by summaries (e.g., $Q_i$, $\sum_j p_{ij} q_{ij}$, campus-level totals, and similar statistics) and modest post-processing in modern data platforms [12].

*2.3 A minimal Simpson example*

I illustrate the ranking problem with a simple two-product, two-campus example. Consider products $A$ and $B$, and campuses $E$("expensive") and $C$("cheap"). Location-specific prices are:

| | $E$ | $C$ |
|---|---|---|
| $A$ | 10 | 4 |
| $B$ | 12 | 6 |

At each campus, product $A$ is strictly cheaper than product $B$: $10 < 12$ at $E$, and $4 < 6$ at $C$.

Now suppose deployment quantities are:

| | $E$ | $C$ | $Q_i$ |
|---|---|---|---|
| $A$ | 90 | 10 | 100 |
| $B$ | 10 | 90 | 100 |

Thus $A$is deployed mostly at the expensive campus, while $B$is deployed mostly at the cheap campus. The naive blended prices are:

$$\bar{p}_A^{(naive)} = 0.9 \cdot 10 + 0.1 \cdot 4 = 9.4,$$
$$\bar{p}_B^{(naive)} = 0.1 \cdot 12 + 0.9 \cdot 6 = 6.6.$$

Despite $A$ being cheaper than $B$ at both locations, the naive operator implies $\bar{p}_A^{(naive)} > \bar{p}_B^{(naive)}$: globally, $A$ appears more expensive. This is a textbook instance of **Simpson's paradox** in a cost aggregation context: aggregating with product-specific location weights reverses an unambiguous local ordering. In large-scale systems with many products and locations, such reversals can be frequent and consequential if blended prices feed into unit cost, pricing, or investment analyses. The goal of the methods developed in the next sections is to construct alternative aggregation operators that (i) preserve total cost (P1), (ii) greatly reduce the scope for such reversals (P2), and (iii) remain simple enough for practical implementation (P3).

## 3. Method 1: Two-Way Fixed-Effect World Prices

This section introduces a first aggregation operator based on a two-way fixed-effect (FE) decomposition of location-specific prices. The central idea is to treat product and campus effects symmetrically: I explain observed prices $p_{ij}$ as the sum of a product-specific "world price" component and a campus-specific cost premium or discount, estimated by weighted least squares. The estimated product effects, after a simple scalar adjustment, define cost-preserving blended prices.

*3.1 Model*

Recall that $p_{ij}$ denotes the observed unit price of product $i$ at campus $j$, and $q_{ij}$ the corresponding quantity. I posit the following additive model:

$$p_{ij} = \alpha_i + \gamma_j + \varepsilon_{ij},$$

where $\alpha_i$ is the product fixed effect, interpreted as the location-adjusted world price of product $i$; $\gamma_j$ is the campus fixed effect, capturing systematic cost level differences across campuses (e.g., due to electricity, cooling, or labor); $\varepsilon_{ij}$ is a residual term. Given the cross-sectional nature of the problem, I estimate this model via quantity-weighted least squares, using the observed deployment quantities $q_{ij}$ as weights. The objective is

$$\min_{\{\alpha_i\},\{\gamma_j\}} \Sigma_i \Sigma_j \, q_{ij} \left(p_{ij} - \alpha_i - \gamma_j\right)^2$$

Because the model is over-parameterized up to a constant shift (adding a constant to all $\alpha_i$ and subtracting it from all $\gamma_j$ yields the same fitted values), I impose a normalization $\sum_{j=1}^{J} \gamma_j = 0$ to ensure identification. Intuitively, this regression "explains" price variation by a product component and a location component, using quantities as importance weights. The estimated product effects $\hat{\alpha}_i$ capture average prices after stripping out systematic campus cost differences; the campus effects $\hat{\gamma}_j$ summarize the relative expensiveness of each campus.

*3.2 Estimator and cost normalization*

Let $\hat{\alpha}_i$ and $\hat{\gamma}_j$ denote the solution to the weighted FE problem. A natural preliminary world price for product $i$ is then $\tilde{p}_i \equiv \hat{\alpha}_i$. However, the FE regression is not explicitly constrained to preserve the exact realized total cost $C = \Sigma_i \Sigma_j \, p_{ij} q_{ij}$. To enforce property P1 (cost preservation), I introduce a scalar adjustment. Define the final blended price as $\bar{p}_i^{(FE)} \equiv \tilde{p}_i + \delta$, where $\delta$ is chosen such that

$$\sum_{i=1}^{I} \bar{p}_i^{(FE)} \, Q_i = C, Q_i = \sum_{j=1}^{J} q_{ij}.$$

Substituting,

$$\sum_i (\tilde{p}_i + \delta) Q_i = \sum_i \tilde{p}_i \, Q_i + \delta \sum_i Q_i = C,$$

so the unique adjustment is $\delta = \frac{C - \sum_i \tilde{p}_i Q_i}{\sum_i Q_i}$. Thus the FE-based world prices are

$$\bar{p}_i^{(FE)} = \hat{\alpha}_i + \frac{C - \sum_k \hat{\alpha}_k \, Q_k}{\sum_k Q_k}.$$

This is a simple one-dimensional correction: I shift all product prices by the same constant to align total cost.

*3.3 Guarantees and robustness boundary*

This subsection summarizes the key guarantees of the two-way FE world-price operator; detailed proofs and edge-case examples are provided in Appendix A.

(P1 Cost preservation). After estimation, I apply a scalar normalization so that the resulting world prices reproduce the realized system expenditure when multiplied by product totals (Appendix A.1).

(P2 Dominance behavior in a balanced matrix). In a fully observed product-campus matrix estimated in levels with standard two-way OLS, the product effect is equivalent to the row mean up to a constant shift; therefore, strict campuswise dominance cannot be reversed by the FE world prices (Appendix A.2, Proposition A1).

Approximate robustness. When prices are close to additive $p_{ij} = \alpha_i + \beta_j + \varepsilon_{ij}$ and residual differences are small relative to cross-product gaps, the FE ranking error is limited and the Ordering Violation Rate (OVR) is correspondingly small (Appendix A.3).

When FE is safe. The FE operator is most reliable when SKU coverage across campuses is broad (near-balanced) and campus premia are approximately parallel across products (weak SKU×campus interactions).

When problems may arise. Sparse coverage, strong SKU×campus interactions, or production practices such as regularization/imputation can yield weak identification and "implicit negative-weight" behavior, which can materially increase OVR/CDR; a minimal failure mode is provided in Appendix A.4.

*3.4 Implementation in distributed systems*

The FE method is straightforward to implement in PySpark or SQL-backed environments and satisfies the implementability requirement (P3). A typical workflow is:

- Step 1. Aggregation. Construct a panel of $(p_{ij}, q_{ij})$ at the product-campus level via a single GROUP BY product_id, campus_id operation.
- Step 2. FE estimation. Export this panel to a regression routine (e.g., Spark MLlib linear regression with dummy variables for products and campuses, using $q_{ij}$ as sample weights) to obtain $\hat{\alpha}_i$ and $\hat{\gamma}_j$.
- Step 3. Cost normalization. Compute $\delta$ using the formula above from the aggregated $\hat{\alpha}_i$ and $Q_i$, and form $\bar{p}_i^{(FE)}$.
- Step 4. Storage and reuse. Write the $\bar{p}_i^{(FE)}$ back as a dimension table keyed by product, so that subsequent cost calculations can be performed by simple joins without re-estimating the model.

All operations beyond the initial regression involve only low-dimensional aggregates and scalar arithmetic. Consequently, the FE operator can be recomputed regularly as prices and deployment patterns evolve, providing an operationally inexpensive and conceptually disciplined alternative to naive blended pricing. This distributed workflow makes the FE operator a key component of a decision support system, enabling real-time integration into DSS pipelines for ongoing monitoring and data-driven operational decisions in cloud computing environments.

## 4. Method 2: Constrained Convex Weight Aggregation

The fixed-effect approach treats product and campus effects symmetrically and recovers world prices as product fixed effects. A complementary strategy is to stay closer to the classical weighted-average intuition, but to choose a single set of campus weights that is (i) common across products, (ii) exactly cost-preserving, and (iii) as close as possible to a chosen baseline weighting scheme. This section develops such an operator as the solution to a small convex optimization problem.

*4.1 Unified campus weights*

I seek a vector of campus weights $w = (w_1, \ldots, w_J)^\top$ satisfying the simplex constraints $w_j \geq 0$, $\sum_{j=1}^{J} w_j = 1$. Given $w$, the blended price for product $i$ is defined as $\bar{p}_i^{(W)} \equiv \sum_{j=1}^{J} w_j\ p_{ij}$. Two features are key:

The weights $w_j$ do not depend on $i$. All products are aggregated with the same location weights. This removes the main driver of Simpson's paradox in the naive operator, namely that each product uses its own location mix as weights.

The blended price is a simple linear combination of observed prices with nonnegative weights. As I emphasize below, this implies that any componentwise dominance relationship across locations is preserved: if product $i$ is no more expensive than product $k$ at every campus, a common nonnegative weight vector cannot reverse their ordering.

To anchor the weights in practice, I introduce a baseline weight vector $\tilde{w} = (\tilde{w}_1, \ldots, \tilde{w}_J)^\top$, representing a preferred notion of campus importance. Typical choices include:

global quantity shares: $\tilde{w}_j = \frac{\sum_i q_{ij}}{\sum_{i,j} q_{ij}}$,

3. or energy-weighted or cost-weighted campus shares derived from auxiliary information.

The constrained convex weight method chooses $w$ as a small adjustment to $\tilde{w}$ that restores exact cost preservation.

*4.2 Cost preservation and optimization problem*

Recall that realized total cost is $C = \Sigma_i \Sigma_j\ p_{ij} q_{ij}$. If blended prices $\left\{\bar{p}_i^{(W)}\right\}$ are to satisfy cost preservation (P1), I require $\sum_{i=1}^{I} \bar{p}_i^{(W)}\ Q_i = \sum_{i=1}^{I} \left(\sum_{j=1}^{J} w_j\ p_{ij}\right) Q_i = C$, with $Q_i = \sum_j q_{ij}$. Rewriting, $\Sigma_i \Sigma_j\ w_i p_{ij} Q_i = \sum_j w_j \left(\sum_i p_{ij}\ Q_i\right) = C$.

Define $A_j \equiv \sum_{i=1}^{I} p_{ij}\ Q_i$, the total cost "exposure" to campus $j$ when computing costs using product-level total quantities $Q_i$. Then the cost-preservation constraint becomes a single linear restriction on $w$: $\sum_{j=1}^{J} w_j\ A_j = C$. I now choose $w$ to be as close as possible to the baseline $\tilde{w}$ while satisfying the simplex and cost constraints. A natural criterion is squared Euclidean distance, leading to the quadratic program

$\min_{w \in \mathbb{R}^J} \sum_{j=1}^{J} (w_j - \tilde{w}_j)^2$, subject to $w_j \geq 0$, $\sum_{j=1}^{J} w_j = 1$, $\sum_{j=1}^{J} w_j\ A_j = C$.

The feasible set is the intersection of the probability simplex with a cost-preservation hyperplane. The objective is strictly convex, so there exists a unique solution $w^\star$. The resulting aggregation operator $\mathcal{A}^{(W)}$ maps $(p_{ij}, q_{ij})$ to world prices

$$\bar{p}_i^{(W)} = \sum_{j=1}^{J} w_j^\star\ p_{ij}.$$

If the baseline weights $\tilde{w}$ already satisfy the cost constraint, they lie in the feasible set; the minimizer is then $w^\star = \tilde{w}$. When the constraint is slightly violated, the solution performs the smallest possible adjustment to $\tilde{w}$ (in squared distance) needed to restore exact cost preservation.

*4.3 Guarantees, solution structure, and computational complexity*

This subsection summarizes the key guarantees of the constrained convex-weight operator and provides an implementation-ready solution procedure. Detailed derivations and proofs are provided in Appendix B.

- P1 Exact system-level cost preservation. By construction, the operator enforces the accounting identity $\sum_i Q_i \tilde{p}_i = C$through a linear constraint on the unified campus weights (Appendix B.1).
- P2 Dominance-robust ranking under ordered campuses. The operator uses a single nonnegative campus weight vector shared across products. When campuswise dominance holds for a product pair $(a, b)$, i.e., $p_{aj} \leq p_{bj}\forall$j, any common nonnegative weighting preserves the ordering $\tilde{p}_a \leq \tilde{p}_b$. Therefore, dominance reversals cannot occur under this operator in the ordered case (Appendix B.1).
- P3 Distributed implementability. The optimization depends only on low-dimensional campus-level summaries; after distributed aggregation, the weight computation is a small post-processing step with negligible runtime for typical $J$(Appendix B.4).
- Closed-form solution without nonnegativity**.** Let $w^0 \in \mathbb{R}^J$be a baseline weight vector. Define the campus exposure vector $s \in \mathbb{R}^J$by $s_j = \sum_i Q_i p_{ij}$and realized cost $C = \sum_{i,j} p_{ij} q_{ij}$. Consider the quadratic objective $\min_w \frac{1}{2} \| w - w^0 \|_2^2$subject to the two equalities $\mathbf{1}^\top w = 1$and $s^\top w = C$, ignoring $w \geq 0$. With $A = \begin{bmatrix} \mathbf{1}^\top \\ s^\top \end{bmatrix} \in \mathbb{R}^{2\times J}, b = \begin{bmatrix} 1 \\ C \end{bmatrix}$, the unique solution is the Euclidean projection
$$w^\star = w^0 - A^\top (AA^\top)^{-1}(Aw^0 - b),$$
where $AA^\top$is $2 \times 2$and thus inverted in constant time (Appendix B.2).
- Nonnegativity via a simple active-set loop. If $w^\star$contains negative entries, I enforce $w \geq 0$using an active-set strategy: fix negative components to zero and re-apply the same closed-form projection on the remaining free indices until all entries are nonnegative. This procedure is summarized in Algorithm 1 and justified in Appendix B.3.

  Algorithm: Active-set projection for unified campus weights

  Input: baseline weights $w^0$, exposure vector $s$, realized cost $C$.

Initialize free set $\mathcal{F} \leftarrow \{1, \dots, J\}$, fixed set $\mathcal{Z} \leftarrow \emptyset$.

Repeat:

(a) Set $w_\mathcal{Z} = 0$. Compute $w_\mathcal{F}$by the closed-form projection using $A_\mathcal{F} = [1_\mathcal{F}^\top; s_\mathcal{F}^\top]$.

(b) If $w_\mathcal{F} \geq 0$, stop. Otherwise move indices with $w_j < 0$from $\mathcal{F}$to $\mathcal{Z}$and continue.

Output $w$, and set world prices $\tilde{p}_i = \sum_j w_j p_{ij}$.

- Complexity. Computing $s_j = \sum_i Q_i p_{ij}$is a distributed aggregation cost $O(IJ)$in the dense case (or linear in the number of observed $(i, j)$pairs). The local post-processing is $O(J)$for the unconstrained projection, and at worst $O(J^2)$with the active-set loop; for typical campus counts $J$in the tens, this step is negligible (Appendix B.4).

*4.4 Implementation in distributed systems*

The convex weight method is designed to be compatible with distributed data pipelines and satisfies the implementability criterion (P3). A minimal implementation involves:

Step 1. campus-level summaries. Compute: (1) the total quantity $\sum_{i,j} q_{ij}$and campus-level totals $\sum_i q_{ij}$(to define $\tilde{w}$, if chosen as global quantity shares); (2) the cost exposure terms $A_j = \sum_i p_{ij} Q_i$, where $Q_i = \sum_j q_{ij}$. These require only GROUP BY aggregations over product and campus.

Step 2. The small quadratic program. Collect $\tilde{w}$ and $A = (A_1, \dots, A_J)$to the driver. Use any standard numerical library (e.g., numpy for a closed-form solution in small dimensions, or cvxpy for clarity) to solve the QP and obtain $w^{\star}$. Since $J$ is small, this step is numerically inexpensive.

Step 3. Broadcast campus weights and compute blended prices. Store $w^{\star}$ as a small table keyed by campus, broadcast it to workers, and compute $\bar{p}_i^{(W)} = \sum_j w_j^{\star} p_{ij}$ via a JOIN on campus followed by GROUP BY product_id and a weighted sum.

Step 4. Persist world prices. Write the resulting $\bar{p}_i^{(W)}$ to a dimension table for downstream use in cost reporting, scenario analysis, or unit-cost calculations.

The heavy lifting is thus confined to simple aggregations and a low-dimensional convex problem. This makes the constrained convex weight operator a practical and robust alternative to naive blended pricing in large multi-campus AI infrastructure environments, especially when cost accounting and ranking stability are both important. By confining computation to aggregations and low-dimensional optimization, the convex operator enhances decision support systems with scalable, robust aggregation for finance and operations teams in large-scale cloud infrastructure.

## 5. Decision Support Workflow and AI Data Center OPEX Case Study

To assess the performance of alternative blended pricing operators, I evaluate them in two complementary settings: stylized campus-level simulations and a realistic AI data center OPEX illustration. Before presenting these empirical results, I describe a practical decision support workflow for operator selection and introduce two simple, model-free evaluation metrics that directly address the core design objectives of ranking stability and cost accuracy. These metrics, order violation rate (OVR) and cost distortion ratio (CDR), can be computed uniformly for any aggregation operator and dataset, enabling transparent comparison among the naive deployment-weighted blend, the fixed-effects (FE) operator, and the constrained convex common-weight operator.

The operators, diagnostic metrics, and structured selection rules together constitute a scalable decision support system (DSS) for robust blended pricing in multi-campus AI data centers. This DSS combines distributed data aggregation, statistical decomposition, constrained optimization, and automated diagnostics to enhance data-driven decision-making under location heterogeneity and uncertainty, supporting operational research and management science applications in cloud computing infrastructure.

### *5.1 A Practical Decision Support Workflow for Operator Selection*

The three operators considered are: (i) the naive deployment-weighted blend (baseline), (ii) the fixed-effects (FE) operator, and (iii) the constrained convex common-weight operator.

- Step 0 (accounting sanity check). Compute the naive blended prices as a baseline and verify the system-level accounting identity (P1). This provides a reference world-price vector that is cost-preserving by construction, but it may suffer from dominance reversals driven by location mix heterogeneity (Simpson-type aggregation).
- Step 1 (diagnose whether a low-rank/additive structure is plausible). Fit the two-way FE model and evaluate out-of-sample fit (e.g., cross-validated RMSE) and the magnitude of residual interactions. If the FE model achieves stable predictive accuracy and residual interactions are weak, the FE-based operator is preferred for its interpretability (decomposition into component and location effects), robustness to mild missingness, and distributed implementability.
- Step 2 (assess whether dominance robustness is a binding requirement). If the intended downstream use is ranking-sensitive (e.g., identifying the most expensive components, budgeting prioritization, chargeback fairness, or what-if planning under reallocation), evaluate the dominance violation rate (or equivalently, the observed violation rate, OVR) under the naive baseline and/or FE blend. When pairwise ordering stability is critical and violations are non-negligible, particularly in the presence of strongly ordered campuses or highly skewed deployment mixes, the convex common-weight operator is recommended as a "safety layer" that enforces a location-robust benchmark while preserving accounting constraints (exactly or within a controlled tolerance).

- Step 3 (choose the operator).

Use FE-based operator when: (i) campus coverage is reasonably balanced; (ii) interaction terms appear weak (additivity approximately holds); (iii) interpretability and decomposition are needed (e.g., separating "intrinsic component expensiveness" from "location premium"); and (iv) missingness can be handled via model-based imputation.

Use convex common-weight operator when: (i) the deployment mix is highly heterogeneous or sparse; (ii) the downstream task is ranking- or fairness-critical; (iii) ordered-campus dominance robustness is treated as a hard requirement; or (iv) a conservative, constraint-driven aggregation is preferred over parametric structure assumptions.

Keep the naive blend only as an accounting baseline or for use cases where ranking stability is irrelevant and the objective is purely to recover aggregate cost with minimal computation.

In production settings, I recommend an "FE-first, convex-guardrail" default policy: deploy the FE-based operator for its interpretability and efficiency, automatically switching to the convex common-weight operator when diagnostics signal (a) substantial dominance violations under the baseline or FE benchmark, or (b) extreme mix heterogeneity threatening ranking fragility. When strict feasibility fails (Section 4.2), the slack-penalized fallback ensures a near-cost-preserving solution with explicit tolerance control. This diagnostics-driven, structured workflow enhances transparency, auditability, and human-centered decision support, making the system readily integrable into finance- and operations-facing DSS pipelines for ongoing monitoring and automated policy triggers.

### *5.2 Evaluation Metrics for Blended Pricing Operators*

Let $\mathcal{A}$ denote a generic blended pricing operator that maps $\{(p_{ij}, q_{ij})\}_{i,j}$ to world prices $\{\bar{p}_i^{\mathcal{A}}\}_{i=1}^{I}$. I focus on two system-level performance measures:

- an order violation rate (OVR), capturing the frequency with which $\mathcal{A}$ reverses unambiguous local price orderings and thus exhibits Simpson-type behavior; and
- a cost distortion ratio (CDR), capturing the discrepancy between total cost computed from blended prices and the realized total cost.

Both metrics can be further decomposed by product group or scenario if desired, but for clarity I define them in their aggregate form.

#### 5.2.1 Order Violation Rate (OVR)

The OVR targets property P2 (location robustness). Intuitively, I examine all product pairs for which one product is strictly cheaper than the other at every campus and ask how often the blended index reverses this ordering.

Consider all unordered product pairs $(i, k)$ with $i \neq k$. For each pair, define the local dominance indicator

$$D_{ik} \equiv \begin{cases} 1, \text{ if } p_{ij} \leq p_{kj}, \forall j \text{ and } p_{ij} < p_{kj} \text{ for some } j, \\ 1, \text{ if } p_{kj} \leq p_{ij}, \forall j \text{ and } p_{kj} < p_{ij} \text{ for some } j \\ 0, \quad \text{otherwise} \end{cases}$$

Thus $D_{ik} = 1$if one of the two products is componentwise cheaper than the other across all campuses, and $D_{ik} = 0$if local prices are mixed and no dominance relation exists.

Given an operator $\mathcal{A}$, define a violation indicator for each dominant pair:

$$V_{ik}^{\mathcal{A}} \equiv \begin{cases} 1, & \text{if } D_{ik} = 1 \text{ and the blended prices reverse the local ordering,} \\ 0, & \text{otherwise.} \end{cases}$$

For example, if $p_{ij} \leq p_{kj}$for all $j$with strict inequality for some $j$(so $i$is locally cheaper everywhere), yet $\bar{p}_i^{\mathcal{A}} > \bar{p}_k^{\mathcal{A}}$, then $V_{ik}^{\mathcal{A}} = 1$.

The order violation rate (OVR) of operator $\mathcal{A}$is defined as

$$\mathrm{OVR}(\mathcal{A}) \equiv \frac{\sum_{i<k} V_{ik}^{\mathcal{A}}}{\sum_{i<k} D_{ik}}.$$

This is the fraction of all locally dominant pairs for which the blended prices contradict the unambiguous campus-level ordering. A lower OVR indicates better location robustness. In particular, by construction of common nonnegative weights, the constrained convex weight operator yields $\mathrm{OVR} = 0$for strictly dominant pairs, whereas the naive operator can exhibit substantial OVR in heterogeneous deployment settings.

5.2.2 Cost Distortion Ratio (CDR)

The CDR targets property P1 (cost preservation). It measures how closely total cost computed using blended prices matches the realized total cost.

For operator $\mathcal{A}$, define the blended total cost

$$C^{\mathcal{A}} \equiv \sum_{i=1}^{I} \bar{p}_i^{\mathcal{A}} \; Q_i, Q_i = \sum_{j=1}^{J} q_{ij},$$

and recall the realized total cost $C = \sum_{i=1}^{I} \sum_{j=1}^{J} p_{ij} \; q_{ij}$.

The cost distortion ratio (CDR) is

$$\mathrm{CDR}(\mathcal{A}) \equiv \frac{|\, C^{\mathcal{A}} - C \,|}{C}.$$

By construction, both the fixed-effect operator and the constrained convex weight operator satisfy $\mathrm{CDR}(\mathcal{A}) = 0$exactly. For naive or ad hoc blended indices that do not enforce cost preservation, $\mathrm{CDR}(\mathcal{A})$provides a simple, scale-free measure of the misalignment between blended-price-based cost and realized cost, which can be nontrivial in the presence of strong location heterogeneity.

I additionally validate the proposed operators using two controlled campus-level simulations (Scenario A and B) designed to stress-test Simpson-type dominance reversals under heterogeneous deployment mixes. In both scenarios, the naive deployment-weighted blend exhibits substantial ranking violations ( $\mathrm{OVR}_{\mathrm{A}}^{Naive} = 0.33, \mathrm{OVR}_{\mathrm{B}}^{Navie} = 0.30$), while the proposed operators eliminate dominance reversals in the ordered cases (OVRs are 0) and materially reduce violations more generally; full designs, parameters, and results are reported in Appendix C. With these diagnostics in place, I next evaluate the operators in a real-world-motivated AI data-center OPEX illustration.

*5.3 Setup and Data Generation for the AI DC OPEX Illustration*

To illustrate the proposed aggregation operators in a more realistic setting, I construct a stylized AI data center (DC) case study. I consider $M = 10$large-scale DC locations, indexed by $m = 1, \dots, 10$, representing anonymized regions in North America and Europe. For each location, I specify an electricity price in the range of approximately 9-18 cents per kWh and a power usage effectiveness (PUE) between 1.2 and 1.6, consistent with recent industry reports on retail/commercial power prices and typical PUE values for enterprise and hyperscale facilities. The effective facility energy price at DC $m$is given by the product of its retail price and PUE.

On the product side, I define $P = 6$ standardized AI compute SKUs, indexed by $p = 1, \ldots, 6$. These SKUs can be interpreted as training-heavy, inference-heavy, or mixed server configurations that are deployed across multiple DCs. For each SKU $p$, I assign a SKU-specific cost multiplier that shifts its operating expense (OPEX) level up or down relative to other SKUs, while keeping all unit costs within a realistic band. I then interpret $p_{pm}$ as the OPEX-based unit cost per normalized compute-hour for SKU $p$ at DC $m$. Concretely, I assume that one normalized compute-hour consumes roughly three kWh at the IT load; multiplying by the local effective energy price and a markup factor for cooling, facilities, and operations yields a unit OPEX in the range of about \$0.6-\$1.4 per compute-hour. This magnitude is broadly consistent with publicly reported GPU cloud prices where energy and operations account for a non-trivial but not dominant share of total hourly cost.

To generate quantities, I model $q_{pm}$ as the annual compute-hours deployed for SKU $p$ at DC $m$. For each SKU, I specify a 10-dimensional deployment pattern across DCs and scale it so that the total annual compute-hours per SKU is on the order of $5 \times 10^5$. Training-oriented SKUs are deliberately concentrated in a subset of higher-cost "AI hub" DCs, while inference-oriented SKUs are more evenly spread and tilted toward lower-cost regions. This design creates heterogeneous deployment mixes across locations and SKUs, mimicking real-world capacity planning decisions where latency, demand, and grid constraints all matter.

The resulting dataset therefore consists of a $P \times M$ panel of unit costs $p_{pm}$ and quantities $q_{pm}$ that are both calibrated to plausible AI DC OPEX levels and intentionally structured to produce Simpson-type reversals under naive blended prices. In the empirical analysis below, I treat each SKU as a "product" and each DC as a "campus" in my notation, and apply the naive, fixed-effect, and convex-weight operators to compute blended world prices and evaluate their ranking performance.

*5.4 Blended prices and ranking in the AI DC case*

I now apply the three aggregation operators to the AI data center OPEX dataset described above. For each SKU $p$, the naive index computes a blended world price by averaging local unit OPEX across data centers using SKU-specific deployment weights, $\bar{p}_p^{\text{naive}} = \sum_m \omega_{pm}^{\text{naive}} p_{pm}$, where $\omega_{pm}^{\text{naive}}$ is the share of annual compute-hours of SKU $p$ hosted at DC $m$. The FE-based index replaces these SKU-specific weights with SKU fixed effects estimated from a two-way regression with DC and SKU effects, followed by a common shift to restore exact cost preservation. The convex-weight index further restricts the weights to be a single set of nonnegative DC weights that minimize the deviation from the global deployment mix while exactly matching total realized OPEX.

Figure 1 and Table 2 summarize the results. Panel (a) plots local unit OPEX for all six SKUs across the ten data centers. By construction, there is substantial cross-location heterogeneity: unit costs in the most expensive hubs (DC1-DC3) are roughly three times those in the cheapest regions (DC8-DC9), reflecting differences in electricity prices and PUE. At the same time, within each data center the ranking of SKUs is stable and interpretable: training-heavy configurations (SKU1-SKU2) are uniformly more expensive than mixed SKUs (SKU3-SKU4), which in turn are more expensive than inference-oriented SKUs (SKU5-SKU6). In other words, the local cost hierarchy is strictly monotone and common across locations.

Panel (b) shows that naive blending severely distorts this hierarchy once deployment patterns are taken into account. Because $\bar{p}_p^{\text{naive}}$ is weighted by each SKU's own deployment mix, its world prices are strongly driven by where workloads currently run. In Table 3, the naive index ranks the inference SKUs SKU5-SKU6 as the two most expensive globally (0.25 and 0.23 USD per compute-hour above the cheapest SKU), even though they are strictly cheaper than all other SKUs in every individual data center. Conversely, training-heavy SKUs that are more heavily deployed in cheaper regions appear artificially inexpensive in world-price space. This manifests as a high order-violation rate, OVR is about 0.73 for the naive index in the AI DC case, despite the smooth and monotone local ordering shown in panel (a).

The proposed operators largely remove this dependence on the realized spatial mix. The FE index fits a two-way model $p_{pm} = \alpha_p + \gamma_m + \varepsilon_{pm}$, interprets the estimated $\alpha_p$(after a common shift) as SKU world prices, and thereby nets out DC fixed effects. Its world prices in Table 3 recover the intended hierarchy (SKU1 > … > SKU6) with OVR essentially zero and negligible cost distortion (CDR on the order of $10^{-16}$). The convex-weight index goes one step further by enforcing common, nonnegative location weights. The resulting world prices again produce a strictly monotone ranking consistent with panel (a), and remain exactly cost-preserving up to numerical precision (CDR ≈ $10^{-11}$).

To summarize, the AI DC illustration shows that, once SKU-location interactions and heterogeneous deployment mixes are present, naive blended prices can generate pervasive Simpson-type reversals in global rankings, even when local costs are completely ordered. The FE and convex-weight indices deliver global cost benchmarks that are tightly aligned with the underlying local OPEX structure, achieving near-zero OVR while maintaining cost recovery, and thus provide a more reliable basis for comparing and managing AI compute SKUs.

**Table 2.** AI data center OPEX illustration (M = 10 data centers, P = 6 AI compute SKUs)

| Product | Naive world price | FE world price | Convex world price |
|---|---|---|---|
| SKU1 | 0.83 | 1.05 | 1.07 |
| SKU2 | 0.77 | 1.01 | 1.02 |
| SKU3 | 0.86 | 0.98 | 0.97 |
| SKU4 | 0.81 | 0.94 | 0.92 |
| SKU5 | 1.25 | 0.88 | 0.87 |
| SKU6 | 1.15 | 0.81 | 0.82 |
| OVR | 0.73 | 0.00 | 0.00 |
| CDR | 0.00 | 0.00 | 0.00 |

*Notes:* World prices are blended unit OPEX per normalized compute-hour for AI compute SKUs SKU1-SKU6 across ten anonymized data center locations, computed under the naive, FE, and convex-weight aggregation operators. Local prices $p_{pm}$are calibrated from stylized electricity prices, PUE values, and SKU-specific cost factors as described in Section 6.1. OVR is the order-violation rate relative to the uniform local cost hierarchy (training-heavy SKUs more expensive than inference-oriented SKUs at every location), and CDR is the cost distortion ratio. Prices are rounded to two decimal places; OVR and CDR are shown with four and up to six decimal places.

**Figure 1.** Local and blended world OPEX for AI compute SKUs

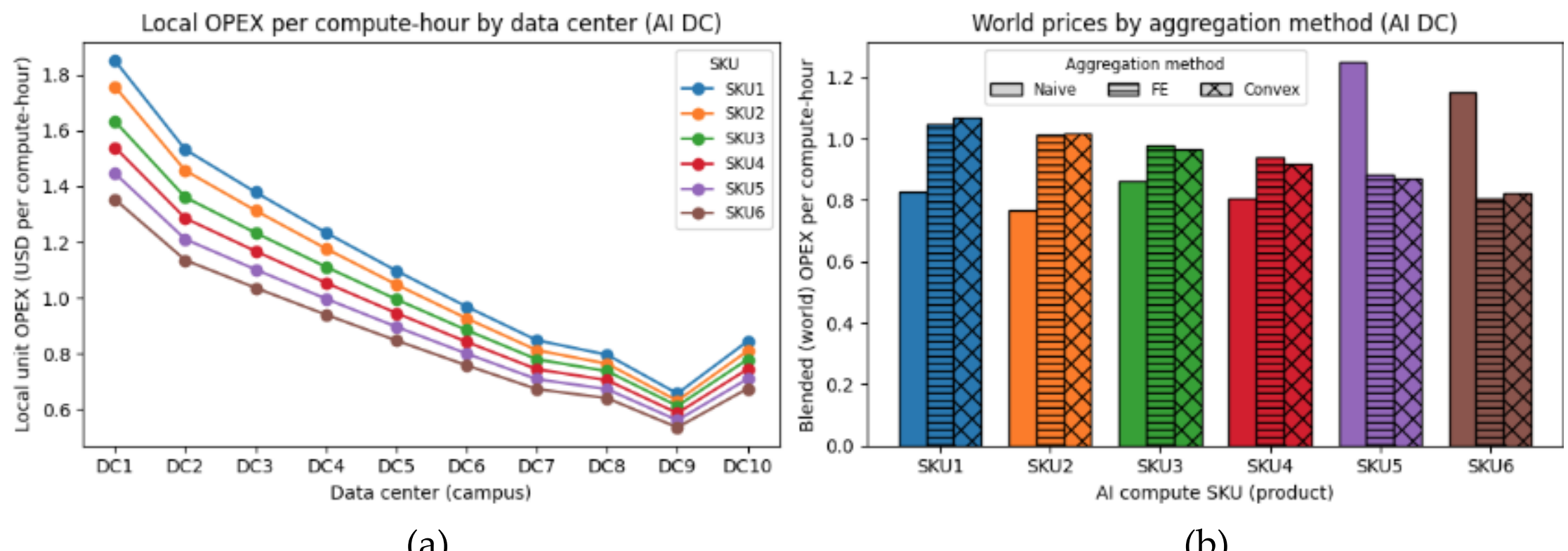

(a) (b)

*Notes:* Panel (a) reports local unit OPEX per normalized compute-hour for six AI compute SKUs (SKU1-SKU6) across ten anonymized data centers DC1-DC10, calibrated from stylized electricity prices and PUE values. Costs vary substantially across locations but exhibit a common local hierarchy, with training-heavy SKUs uniformly more expensive than mixed and inference-oriented SKUs. Panel (b) shows the corresponding blended "world" prices under the naive, fixed-effects (FE), and convex-weight aggregation operators. Naive world prices are computed using SKU-specific deployment weights and therefore reflect current workload geography, whereas the FE and convex-weight indices strip out campus effects and impose common nonnegative location weights while preserving total realized OPEX.

*5.5 Interpretation for AI capacity planning*

For AI infrastructure teams, the patterns in Table 3 and Figure Z have direct operational consequences. In practice, finance and capacity-planning groups almost always compress the underlying multi-location cost structure into a single global unit cost per SKU. That scalar is then used for internal chargeback to product teams, to evaluate scenarios such as "shift 20% of training from DC3 to DC9," and to benchmark list prices or discounts in cloud offerings. When this global benchmark is computed as a naive blend over the current deployment mix, it effectively hard-codes today's geography into tomorrow's decisions: a temporary concentration of workloads in a high-cost hub can make a SKU look intrinsically expensive, while SKUs with more exposure to low-cost regions appear artificially cheap. In the AI DC illustration, this shows up as high OVR for the naive index in Table 3 and as visibly distorted spreads in the right-hand panel of Figure Z, even though panel (a) reveals a smooth and monotone local hierarchy.

The proposed FE and convex-weight indices offer a more decision-robust alternative. Both operators preserve total realized OPEX to machine precision, so aggregate budget constraints and cost recovery analyses remain unchanged. At the same time, by stripping out campus fixed effects and enforcing common, nonnegative location weights, they deliver location-robust world prices with essentially zero OVR. For planners, this means that global unit costs are driven by the underlying technology and efficiency of SKUs rather than by incidental siting choices. Such indices are therefore more appropriate as baselines for long-horizon capacity roadmaps, region-expansion studies, and "what-if" simulations of workload rebalancing, and they provide a cleaner signal for pricing and discount strategies than naive deployment-weighted averages. These robust world prices form a foundational element of a decision support system, delivering location-independent signals that improve the reliability of data-driven decisions in AI cloud computing operations.

## 6. Robustness under Stress Scenarios

This section probes whether the proposed blended-pricing operators remain reliable when the clean assumptions used to motivate them are intentionally strained. This is necessary because Simpson-type reversals are not a "corner case" that disappears after a single illustrative example; they arise from a small number of structural frictions that routinely co-exist in distributed procurement and multi-campus AI operations, namely, heterogeneous deployment mixes, location-driven price dispersion, non-additive SKU×campus effects, and incomplete price coverage. I therefore organize robustness evidence around three stress tests, each designed to isolate one such friction and trace how the ranking signal and cost recovery behave as the friction intensifies. Rather than relying on one synthetic scenario, these tests map out the operators' operational boundaries: when naive SKU-specific aggregation becomes unstable, when an additive two-way FE world-price benchmark remains adequate, and when a constraint-based common-weight operator provides a safer fallback. The goal is not to "win" every stress setting, but to demonstrate a coherent robustness story, showing that the proposed cost-preserving operators systematically reduce ranking reversals and stabilize cross-campus comparisons in precisely the regimes where naive blending is most vulnerable. The goal is not to "win" every stress setting, but to demonstrate a coherent robustness story, showing that the proposed cost-preserving operators systematically reduce ranking reversals and stabilize cross-campus comparisons in precisely the regimes where naive blending is most vulnerable, thereby supporting resilient decision-making under uncertainty in DSS applications.

*6.1 Design and Main Finding: Mix Extremity*

This subsection conducts a mechanism-driven stress test that isolates the canonical failure mode behind Simpson-type reversals in blended pricing, heterogeneous, product-specific deployment mixes interacting with location-dependent prices. I consider the minimal setting with two products $(i \in \{A, B\})$and four campuses $(j = 1, \ldots, 4)$. Local prices satisfy strict campuswise dominance by construction: $p_{Aj} < p_{Bj} \forall j$, so any reversal of the world-price ranking $\tilde{p}_A \leq \tilde{p}_B$must be attributable to the aggregation operator rather than the local price data.

Stress mechanism (controlled mix mismatch). Holding product totals $Q_A, Q_B$fixed, I generate campus-level deployments using an extremity parameter $\eta \in [0,1]$that progressively swaps where each product is deployed. Specifically, product $A$is shifted toward the most expensive campus while product $B$is shifted toward the cheapest campus: $w_A(\eta) = [1 - \eta, 0, 0, \eta]$, $w_B(\eta) = [\eta, 0, 0, 1 - \eta]$, and quantities are set as $q_{ij}(\eta) = Q_i\, w_i(\eta)$. This design deliberately maximizes the potential for Simpson-type reversals under product-specific weighting, because the high-cost campus receives a growing share of the deployment mass for the locally cheaper product.

Evaluation statistic. For each $\eta$, I compute world prices using (i) the naive product-specific blend, (ii) the two-way FE operator with exact cost-preserving normalization, and (iii) the constrained convex common-weight operator. I summarize ranking behavior through $\Delta(\eta) = \tilde{p}_A(\eta) - \tilde{p}_B(\eta)$, where $\Delta(\eta) > 0$constitutes a dominance reversal (the locally cheaper product is ranked as more expensive after aggregation).

Main finding. Figure 2 plots $\Delta(\eta)$over the full extremity path. The naive product-specific blend exhibits a sharp monotone increase in $\Delta(\eta)$and crosses zero once mix mismatch becomes sufficiently extreme, producing dominance reversal for $\eta \gtrsim 0.545$. This provides a transparent demonstration that Simpson-type reversals can arise even when local prices are perfectly ordered by campuswise dominance; the reversal is purely induced by mixing each product with its own location distribution.

**Figure 2.** Mix Extremity Stress Test: Dominance Reversal Path as Deployment Mix Shifts

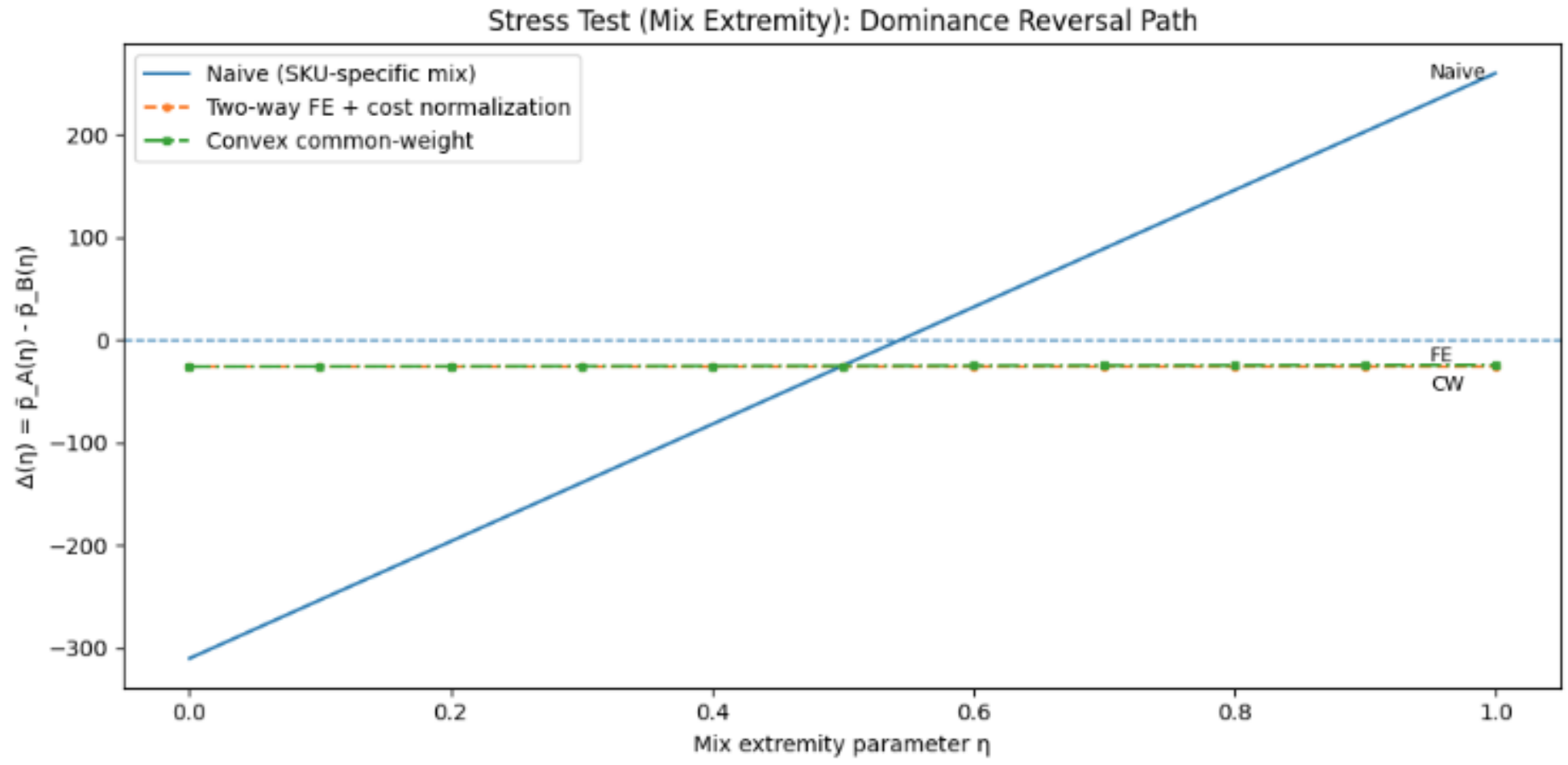

In contrast, both proposed operators remain strictly below zero across the entire $\eta$range, i.e., they preserve the correct ordering $\tilde{p}_A(\eta) \leq \tilde{p}_B(\eta)$under the most adversarial deployment mismatch. This behavior is consistent with their structural safeguards: the convex common-weight operator enforces a single nonnegative campus weighting shared across products, while the two-way FE operator, under a fully observed price matrix in levels, reduces to a row-mean-based ordering up to a common shift, which cannot invert campuswise dominance. Thus, this stress test confirms that the proposed operators eliminate the canonical Simpson failure mode induced by product-specific mixing, even under extreme deployment heterogeneity.

*6.2 Interaction Stress: Boundary of Additive Two-Way FE*

This subsection stress-tests the aggregation operators under SKU×campus interaction, a setting in which local prices deviate from the additive two-way structure that motivates the FE operator. I generate a fully observed price matrix with interaction strength $\gamma$through $\log p_{ij} = \alpha_i + \beta_j + \gamma\,(u_i v_j)$, where $\alpha_i$and $\beta_j$are product and campus components and $u_i v_j$induces non-parallel campus premia across SKUs. Holding total quantities $Q_i$fixed, I impose a fixed mix mismatch across products (a moderate but persistent deployment imbalance across campuses) so that any ranking instability is attributable to the interaction between heterogeneous mixes and interaction-driven location effects, rather than changing product scale.

FE misspecification diagnostic. Figure 3(a) reports the root-mean-square (RMS) residual from the best additive two-way fit in levels, $p_{ij} \approx a_i + b_j$. The RMS residual increases steeply with $\gamma$, confirming that stronger SKU×campus interactions progressively violate the additive structure underlying the FE operator. This diagnostic provides a transparent boundary condition: as interactions intensify, a two-way additive representation becomes increasingly inaccurate in levels.

Ranking implications. Figures 3(b)-(c) summarize how interaction strength affects the world-price ranking signal $\Delta(\gamma) = \tilde{p}_A(\gamma) - \tilde{p}_B(\gamma)$. The naive SKU-specific blend exhibits a reversal for $\gamma \in (0,0.5)$, despite the fact that the stress design holds totals fixed and varies only the interaction-driven non-parallel campus premia. In contrast, the FE-based cost-preserving operator preserves the ordering over the full $\gamma$range in this fully observed setting. Figure 3(c) further shows that the gap $\Delta_{\text{naive}}(\gamma) - \Delta_{\text{FE}}(\gamma)$remains positive and sizable throughout, indicating a systematic upward distortion of the naive ranking signal relative to FE when product-specific mixes interact with campus-specific price responses.

**Figure 3.** Sparsity Stress Test: Ranking Stability Under Missing SKU×Campus Prices

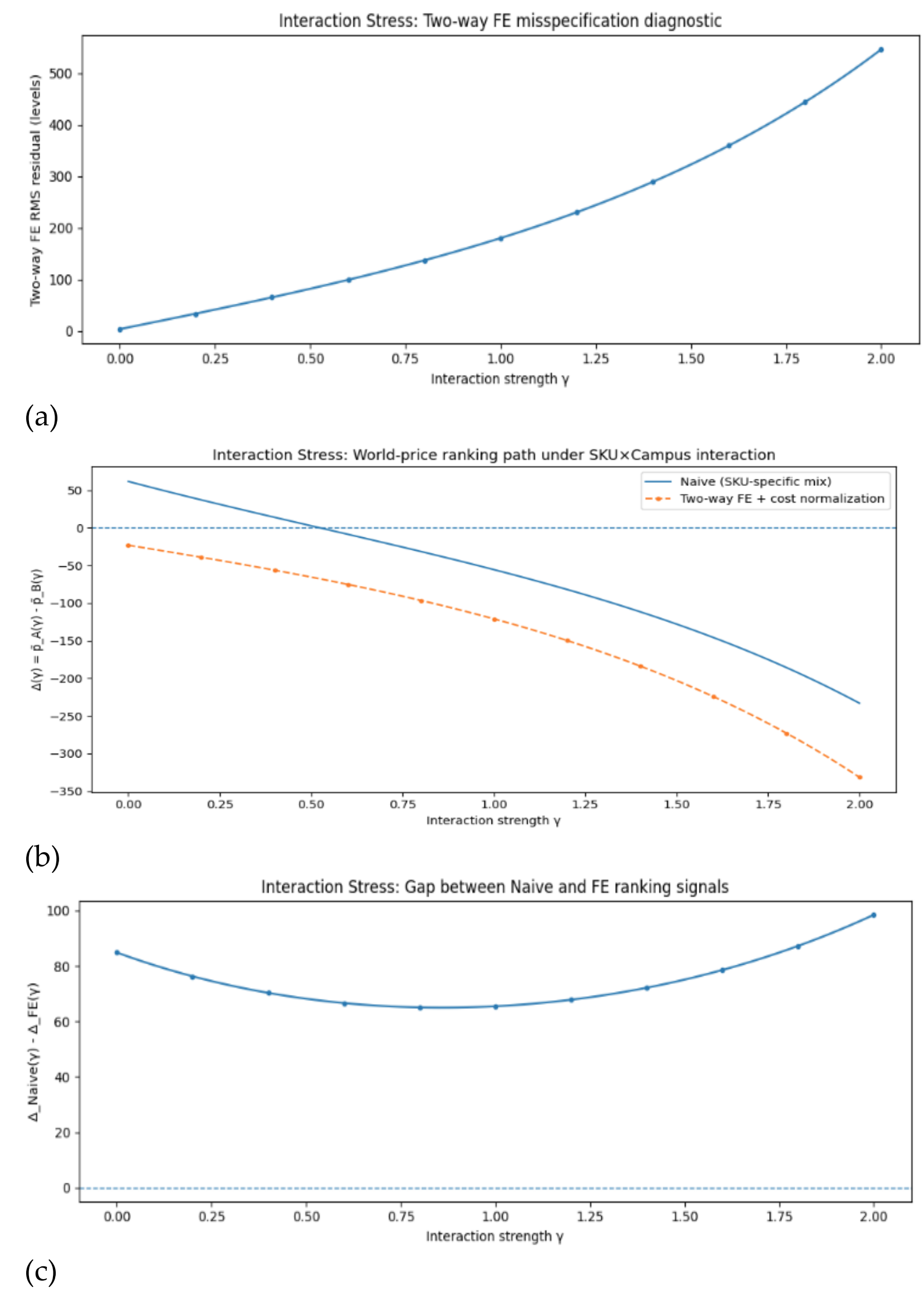

(a)

(b)

(c)

To sum up, this stress test isolates a key mechanism beyond simple mix extremity: non-additive SKU×campus interactions can induce ranking reversals under naive product-specific blending, and the distortion grows as interaction strength increases. The residual diagnostic clarifies that rising interaction strength signals increasing deviation from additivity in levels; while FE remains stable here under full observation, the same diagnostic motivates stronger safeguards in settings with incomplete price matrices or when extrapolation is required, precisely the regimes where the constrained convex operator and additional stress tests are most informative.

*6.3 Sparsity Stress: Missing SKU×Campus Prices*

This subsection evaluates operator robustness when the SKU×campus price matrix is incomplete, a common operational regime in which certain products are not procured, deployed, or recorded at particular campuses. I introduce sparsity by independently masking each price cell $(i, j)$ with probability $\rho \in [0,0.75]$ while enforcing minimal identifiability (each SKU is observed at least twice and each campus has at least one observed SKU). Total quantities $Q_i$, the product-specific deployment mixes $w_{ij}$, and the aggregate accounting cost $C$ are held fixed. For the FE operator, missing prices are imputed using the fitted two-way additive structure in levels; the convex common-weight operator is computed from the implied campus bundle costs after imputation, and both operators apply cost preservation as defined in Section 2.2. Performance is evaluated relative to an oracle benchmark constructed from the full matrix using the cost-preserving common-weight operator, yielding a reference ordering and $\Delta^{\star} = p_A^{\star} - p_B^{\star}$.

Figure 4(a) reports the reversal rate of the ranking signal $\Delta(\rho) = \tilde{p}_A(\rho) - \tilde{p}_B(\rho)$relative to the oracle sign. The naive SKU-specific blend remains highly unstable: even at $\rho = 0$it exhibits near-certain reversals, reflecting that its primary failure mode is the interaction between heterogeneous product mixes and location heterogeneity rather than data incompleteness per se. As $\rho$increases, naive reversal rates remain substantial. In contrast, both proposed operators remain essentially reversal-free across a wide sparsity range, with only a small increase in flips for the convex common-weight operator at very high missingness.

Figure 4(b) reports the MAE of $\Delta(\rho)$relative to the oracle. The naive blend exhibits large, persistent errors, indicating severe distortion not only in sign but also in magnitude. The FE and convex operators reduce $\Delta$error by an order of magnitude, and the convex operator degrades smoothly as sparsity increases, consistent with increased uncertainty in estimating campus bundle costs when a larger share of cell prices must be inferred.

Because observed-cell residuals are not comparable across $\rho$(the evaluated set changes with missingness), Figure 4(c) reports a comparable diagnostic: the FE imputation RMSE on missing cells. The imputation error increases monotonically with $\rho$, confirming that sparsity primarily harms the FE operator through extrapolation, when fewer cells are observed, the additive structure becomes progressively data-starved and imputations $\hat{p}_{ij} = \hat{\mu} + \hat{a}_i + \hat{b}_j$become less accurate. This diagnostic clarifies the operational boundary of the FE approach under incomplete matrices and motivates the convex common-weight operator as a complementary safeguard in sparse, distributed pricing environments.

**Figure 4.** Sparsity Stress Test Under Missing SKU×Campus Prices: Ranking, MAE, and FE Imputation RMSE

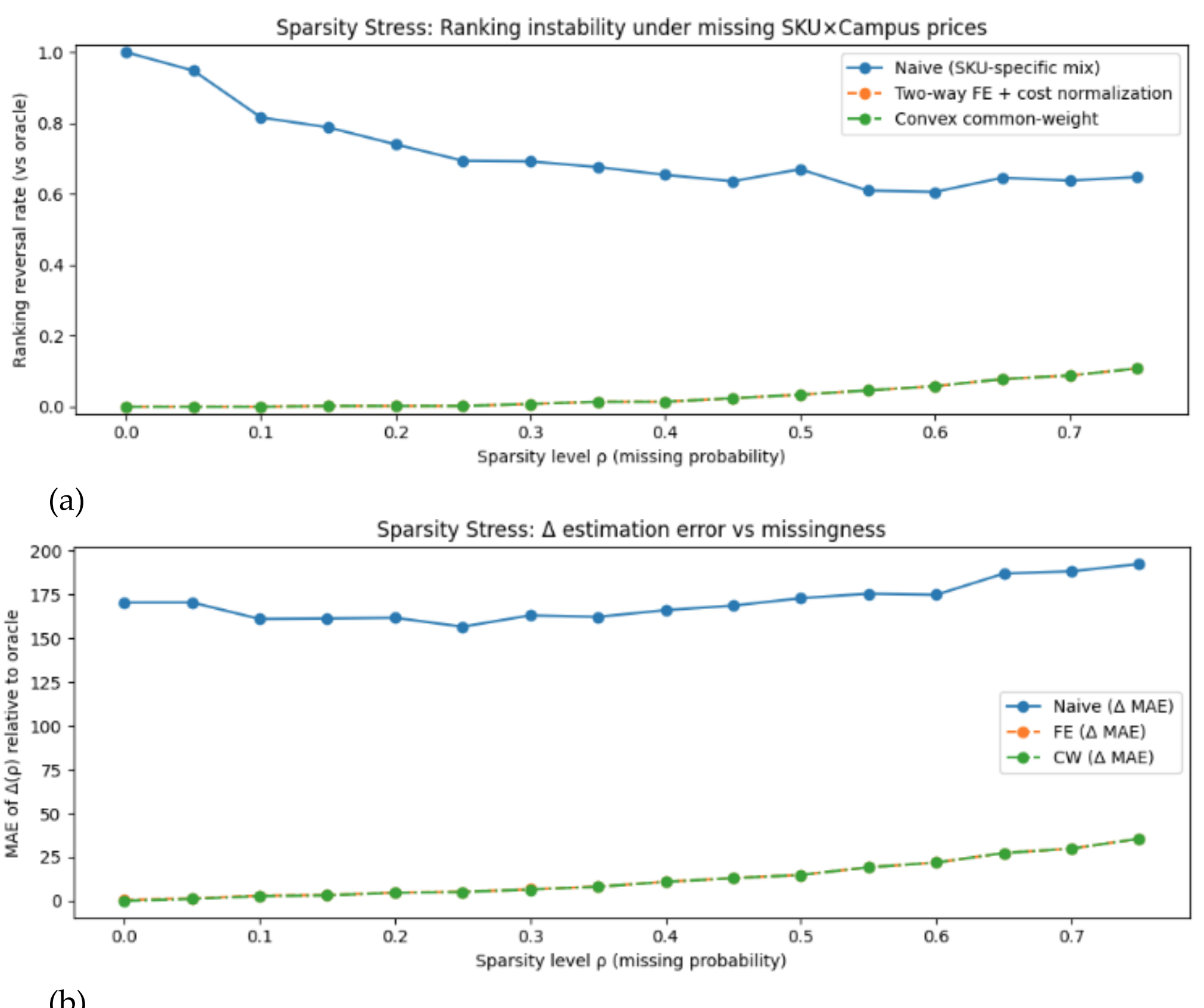

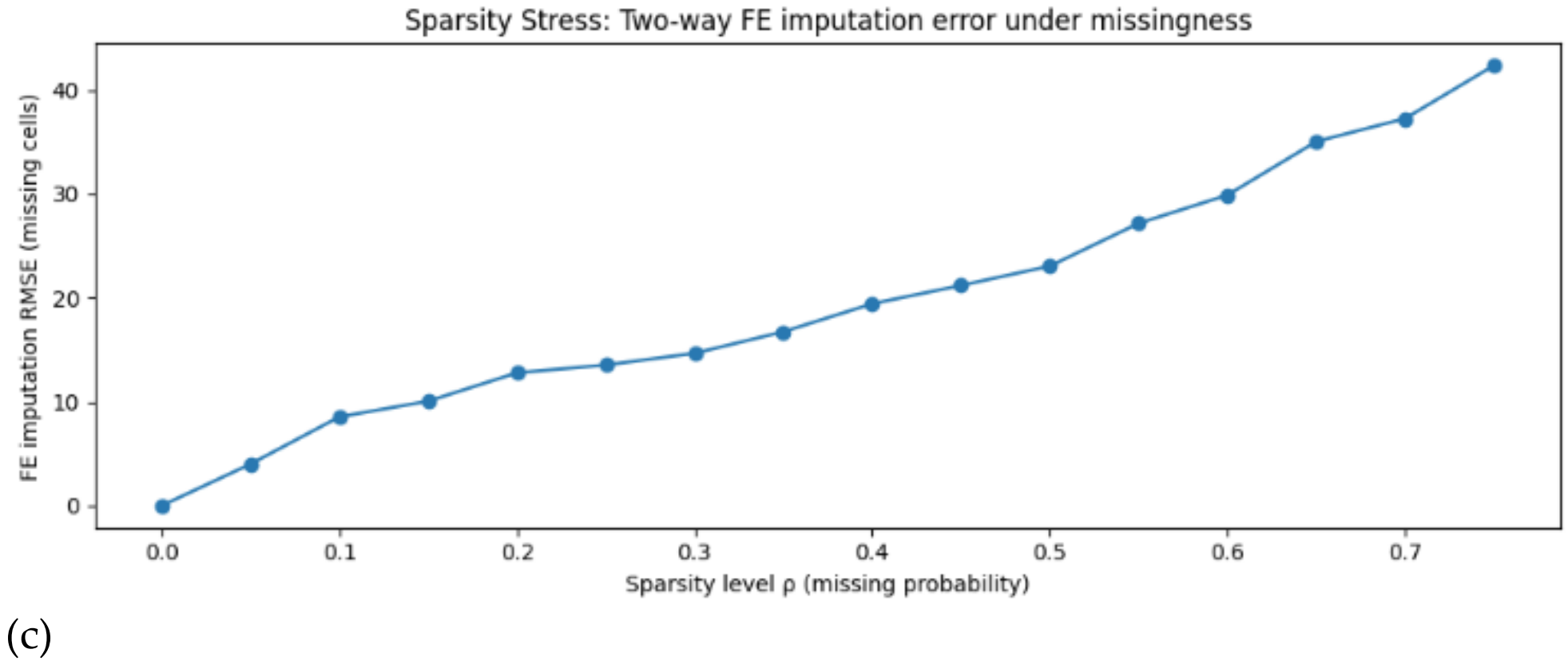

(c)

Thus, the sparsity stress test shows that (i) naive SKU-specific blending is structurally unstable even under full observation, and (ii) the proposed cost-preserving operators deliver robust rankings under missingness, with the FE imputation RMSE providing a transparent measure of when sparsity begins to erode reliability.

## 7. Discussion and Conclusion

This paper addresses a critical challenge in decision-making and decision support systems (DSSs) for large-scale cloud and AI infrastructure: constructing cost-preserving "world prices" for identical SKUs across geographically heterogeneous data center campuses. From a DSS perspective, the task is to transform complex, multi-location cost data into stable, auditable, and actionable benchmarks that support ranking-sensitive decisions in procurement, budgeting, chargeback, capacity planning, and workload reallocation. While naive deployment-weighted blending satisfies accounting constraints by preserving aggregate cost, it remains vulnerable to Simpson-type dominance reversals induced by heterogeneous location mixes, potentially distorting decision signals and leading to suboptimal managerial outcomes. By drawing on the index-number tradition, where aggregation must balance accounting identity with economic meaningfulness, I formalized key requirements for blended pricing operators and evaluated common approaches against them.

The core contribution is a scalable decision support system grounded in operational research and management science. It comprises two complementary operators, diagnostic metrics, and a structured selection workflow. The two-way fixed-effects (FE) operator decomposes prices into interpretable global SKU effects and campus premia, followed by scalar normalization to ensure exact cost preservation; it delivers explainable benchmarks and naturally handles mild missingness through model-based smoothing. The constrained convex common-weight operator, solved via quadratic programming, enforces a unified nonnegative campus weighting scheme that eliminates dominance reversals by construction under ordered-campus conditions, while incorporating feasibility diagnostics and a slack-based fallback for robustness under extreme heterogeneity. Together with model-free evaluation metrics, order violation rate (OVR) and cost distortion ratio (CDR), and the detect-correct-validate workflow, these components provide practitioners with a transparent, diagnostics-driven toolkit (cf. explainable AI–based risk diagnostics in applied decision systems [22]) for trading off interpretability, conservatism, and ranking stability in production environments.

The simulation experiments and realistic AI data center OPEX case study demonstrate that the proposed DSS substantially reduces ranking violations relative to naive blending without sacrificing cost accuracy, delivering more reliable inputs for data-driven decision-making precisely when power, cooling, and infrastructure costs dominate AI-era operations.

In the broader context of AI infrastructure measurement, He (2025) proposes a unified metric architecture linking economic, performance, and efficiency indicators across layers [16]. The present work complements this by supplying robust, cost-preserving aggregation operators that mitigate Simpson-type reversals, thereby enabling decision-ready benchmarking across heterogeneous campuses. The methods are fully compatible with distributed computation and integrate seamlessly into scalable data processing pipelines commonly used in cloud computing environments.

Natural extensions include: (i) the dual problem of inferring campus-level resource prices; (ii) dynamic multi-period settings with time-series smoothing and regime detection; (iii) stochastic or robust formulations to explicitly address price and quantity uncertainty; and (iv) enhanced human-explainable AI (XAI) features, such as interactive visualization of campus premia and violation diagnostics, to further strengthen human-centered DSS design. Although motivated by AI data centers, the framework applies equally to multi-region cost aggregation in energy systems, healthcare, and public infrastructure, domains where location-robust, cost-preserving indicators are essential for effective decision support, including grid-resilient siting via modular landfill remediation [19].

Ultimately, this study advances decision-making processes in applied sciences by delivering an implementable, explainable decision support system that enhances the reliability and fairness of cost benchmarking in large-scale, geographically distributed cloud computing operations.

**Acknowledgments:** The views expressed in this paper are solely those of the author and do not necessarily reflect the views of Google, or any other current or former employer or affiliated institution.

## Appendix A - Theoretical Guarantees and Robustness Boundary of the Two-Way FE Operator

*Appendix A.1 Proof of exact cost preservation via scalar normalization.*

(Here I show that choosing $\kappa$ such that $\sum_i(\hat{\alpha}_i + \kappa)Q_i = C$ enforces P1 exactly, and that $\kappa$ does not affect rankings.)

*Appendix A.2 Dominance* preservation in a balanced matrix.

Proposition A1 (Dominance preservation under balanced two-way OLS in levels).
Consider a fully observed $I \times J$ price matrix and the additive model $p_{ij} = \alpha_i + \beta_j + \varepsilon_{ij}$ estimated by OLS with standard normalization (e.g., $\sum_i \alpha_i = 0$, $\sum_j \beta_j = 0$). Define the FE world price as $\tilde{p}_i = \hat{\alpha}_i + \kappa$, where $\kappa$ is chosen to satisfy P1. If product $a$ is weakly cheaper than product $b$ at every campus, $p_{aj} \leq p_{bj}$, $\forall$ j, then $\tilde{p}_a \leq \tilde{p}_b$. If the dominance is strict in at least one campus, then $\tilde{p}_a < \tilde{p}_b$.
Proof sketch. In a balanced two-way FE, $\hat{\alpha}_i$ equals the row mean of $p_{ij}$ plus a constant independent of $i$. Campuswise dominance implies the row mean of $a$ does not exceed that of $b$. The scalar shift $\kappa$ preserves ordering.

*Appendix A.3 Approximate robustness under near-additivity.*

Corollary A1 (Gap condition for strict ordering).
Suppose $p_{ij} = \alpha_i + \beta_j + \varepsilon_{ij}$ on a balanced matrix and $|\varepsilon_{ij}| \leq \delta$. If $b$ dominates $a$ with uniform margin $\min_j(p_{bj} - p_{aj}) \geq 2\delta$, then $\tilde{p}_a < \tilde{p}_b$. More generally, when $\mathbb{E}[\varepsilon_{ij} \mid i] = 0$ and residuals are independent of deployment mixes, the FE ranking error is controlled by the aggregation of residual differences and is small when residual variation is small relative to cross-product price gaps.

*Appendix A.4 Minimal failure mode outside the balanced additive regime.*

Example A1 (Interaction + sparsity induces ranking instability).
In practice, not every SKU is priced in every campus, and campus premia may be non-parallel across SKUs (SKU×campus interaction). Consider a setting where true prices satisfy $p_{aj} \leq p_{bj}$ across all campuses, but only a sparse subset of $(i, j)$ pairs is observed and missing prices are imputed by an additive FE fit. If the true data exhibit strong SKU×campus interactions, an additive imputation can distort the unobserved entries in a way that breaks the implied dominance pattern and can change the resulting world-price ranking after normalization. This is not a contradiction to Proposition A1 (which requires a fully observed balanced matrix), but it is the practical boundary that motivates reporting OVR/CDR and preferring the common-weight operator when dominance-robustness is critical.

## Appendix B - Guarantees and Solution Details for the Constrained Convex-Weight Operator

*Appendix B.1 Cost preservation and dominance guarantee*

Lemma B1 (Cost preservation under unified weights). Let $Q_i = \sum_j q_{ij}$, $s_j = \sum_i Q_i\, p_{ij}$, and define world prices by $\tilde{p}_i = \sum_j w_j\, p_{ij}$ with a unified weight vector $w$ satisfying $\mathbf{1}^\top w = 1$ and $s^\top w = C$, where $C = \sum_{i,j} p_{ij}\, q_{ij}$. Then the system-level cost implied by world prices equals realized cost:

$$\sum_i Q_i\, \tilde{p}_i \;=\; \sum_i Q_i \sum_j w_j\, p_{ij} \;=\; \sum_j w_j \sum_i Q_i\, p_{ij} \;=\; s^\top w \;=\; C.$$

This establishes (P1) for any feasible $w$.

Proposition B1 (No dominance reversal under common nonnegative weights). Suppose for a product pair $(a, b)$ that $p_{aj} \le p_{bj}$ for all campuses $j$. If $w \ge 0$, then the corresponding world prices satisfy $\tilde{p}_a \le \tilde{p}_b$:

$$\tilde{p}_a - \tilde{p}_b \ = \ \sum_j w_j \, (p_{aj} - p_{bj}) \ \le \ 0.$$

If dominance is strict in at least one campus with positive weight, then $\tilde{p}_a < \tilde{p}_b$. This establishes (P2) in the dominance case.

*Appendix B.2 Closed-form projection without nonnegativity*

Consider the optimization

$$\min_w \frac{1}{2} \| w - w^0 \|_2^2 \ \text{s.t.} \mathbf{1}^\top w = 1, s.t. \ s^\top w = C,$$

ignoring $w \ge 0$. The Lagrangian is

$$\mathcal{L}(w, \lambda) = \frac{1}{2} \| w - w^0 \|_2^2 + \lambda^\top (Aw - b), A = \begin{bmatrix} \mathbf{1}^\top \\ s^\top \end{bmatrix}, b = \begin{bmatrix} 1 \\ C \end{bmatrix}.$$

First-order optimality yields $w - w^0 + A^\top \lambda = 0$, so $w = w^0 - A^\top \lambda$. Substituting into the constraints gives

$$A(w^0 - A^\top \lambda) = b \Rightarrow AA^\top \lambda = Aw^0 - b,$$

hence $\lambda = (AA^\top)^{-1}(Aw^0 - b)$ and

$$w^\star = w^0 - A^\top (AA^\top)^{-1}(Aw^0 - b),$$

which is the Euclidean projection of $w^0$ onto the affine set $\{w : Aw = b\}$. Since $AA^\top \in \mathbb{R}^{2\times 2}$, the inversion is constant-time.

*Appendix B.3 Enforcing nonnegativity via an active-set projection*

When the nonnegativity constraint $w \ge 0$ is imposed, the problem becomes a strictly convex quadratic program with a unique optimum whenever the feasible set is nonempty. The active-set loop used in Algorithm 1 is a standard approach for projecting onto an affine set intersected with the nonnegative orthant: each iteration fixes violated components to zero and recomputes the projection on the remaining free variables. Because at least one index is moved from the free set to the fixed set whenever a violation is found, the algorithm terminates in at most $J$ iterations.

*Appendix B.4 Complexity and practical notes*

The dominant cost is the distributed aggregation of the exposure vector $s$, which is $O(IJ)$ for a dense price matrix and linear in the number of observed $(i, j)$ pairs for sparse inputs. The post-processing step is $O(J)$ for the unconstrained projection and $O(J^2)$ in the worst case with the active-set loop. In typical multi-campus deployments where $J$ is on the order of tens, this step is negligible relative to distributed aggregation. If $s$ is nearly collinear with $\mathbf{1}$ (degenerate case where campuses are nearly identical under exposure), numerical stability can be improved by rescaling $s$ or adding a small ridge term to the $2 \times 2$ system; this does not change feasibility but can improve conditioning.

*Appendix B.5. Feasibility and Robust Fallback*

Feasibility check. The constraint set in (B.1) is nonempty if and only if the target system-level exposure $C$ lies in the convex hull of $\{E_j\}_{j=1}^{J}$; since $w \in \Delta_J$, this reduces to the scalar condition $C \in [\min_j E_j, \max_j E_j]$.

Fallback (soft cost-preservation). When $C \notin [\min_j E_j, \max_j E_j]$ due to missing quantities, coverage gaps, or extreme deployment mix, I replace the hard equality $\sum_j w_j E_j = C$ with a penalized slack:

$$\min_{w \in \Delta_J} \frac{1}{2} \parallel w - u \parallel_2^2 + \frac{\rho}{2}\left(\sum_{j=1}^{J} w_j E_j - C\right)^2,$$

where $\rho > 0$controls the strength of cost preservation. This relaxation always yields a feasible solution, and recovers the exact-constraint optimizer as $\rho \to \infty$. In implementation, $\rho$can be selected to ensure $\mid \sum_j w_j E_j - C \mid \leq \varepsilon$for a predefined tolerance $\varepsilon$(e.g., based on measurement error in $E_j$).

Fallback (boundary projection; optional). Alternatively, if strict feasibility is required, one may project $C$to the closest feasible boundary $\tilde{C} = \text{clip}(C, \min_j E_j, \max_j E_j)$and solve the original constrained QP with $C$replaced by $\tilde{C}$, which minimizes the deviation from the intended accounting target while preserving the simplex constraint.

## Appendix C - Campus-level Simulation Results

*C.1 Simulation setup and goals*

This appendix reports two controlled campus-level simulations (Scenario A and Scenario B) used as additional validation of the blended pricing operators. The simulations are intentionally stylized: they are not intended to reproduce any specific firm's procurement contracts or operational footprint. Instead, their role is to provide a mechanism check under fully controlled conditions, isolating how heterogeneous campus mixes interact with multi-campus price matrices to create (or eliminate) Simpson-type ranking reversals in world-price comparisons.

I construct synthetic price and quantity panels $\{p_{ij}, q_{ij}\}$over products $i = 1, \ldots, I$and campuses $j = 1, \ldots, J$. For each scenario, I specify (i) a campus-level price structure and (ii) product-specific deployment mixes, then compute world prices under (a) the naive deployment-weighted blend and (b) the proposed operators. Performance is evaluated using the ranking-stability diagnostics defined in Section 5, primarily the Ordering Violation Rate (OVR) and Cost Dominance Reversal (CDR) rate. These metrics are designed to detect (i) general ordering inconsistencies in pairwise comparisons and (ii) the specific failure mode where campuswise dominance is reversed after aggregation.

The two scenarios target complementary stress cases. Scenario A is designed to be a clean dominance benchmark: it enforces a common campus ordering in local prices while allowing deployment mixes to differ sharply across products. This setting isolates the classic Simpson mechanism in the most transparent way and serves as a direct test of dominance robustness. Scenario B introduces additional realism by weakening idealized assumptions (e.g., allowing non-parallel campus premia and/or controlled noise and heterogeneity), creating a more challenging environment in which dominance may not hold universally.

Across both scenarios, the goal is not to "win" on a single synthetic instance, but to demonstrate that (i) naive product-specific mixing can generate large OVR/CDR even under seemingly benign campuswise price orderings, and (ii) enforcing a common-weight structure and an accounting identity can eliminate dominance reversals in ordered cases and materially reduce ranking violations more generally.

*C.2 Scenario A: dominance-consistent local prices + heterogeneous mixes*

Figure C1 and Table C1 report results for a fleet with three standardized trays A-C deployed across four campuses E1-E4. Panel (a) of Figure 2 shows that the local ordering is strictly monotone at every campus, with $A < B < C$. Under naive aggregation, however, the global ranking is inverted for A and B: Table 1 shows a naive world price of 9.10 for A versus 6.90 for B, even though A is cheaper than B in every location. This textbook Simpson reversal is reflected in an OVR of 0.33 for the naive index. In contrast, the FE and convex-weight operators recover a monotone world-price vector with $A < B < C$, and both achieve exact cost preservation; their OVR is zero and their CDR is numerically zero.

**Table C1.** Campus-level simulation, scenario A (J = 4 campuses, I = 3 products)

| Product | Naive world price | FE world price | Convex world price |
|---|---|---|---|
| A | 9.10 | 7.00 | 7.00 |
| B | 6.90 | 9.00 | 9.00 |
| C | 11.00 | 11.00 | 11.00 |
| OVR | 0.33 | 0.00 | 0.00 |
| CDR | 0.00 | 0.00 | 0.00 |

*Notes:* World prices are blended "world" prices for products A-C under the naive, fixed-effects (FE), and convex-weight aggregation operators. All campus-specific prices share the same local ordering $A < B < C$; heterogeneity arises only from differences in quantity mixes across campuses. OVR denotes the order-violation rate with respect to this common local ordering, and CDR denotes the cost distortion ratio as defined in Section 5.1. World prices are rounded to two decimal places; OVR and CDR are reported with four and six decimal places, respectively.

**Figure C1.** Campus-level local prices and blended world prices in scenario A

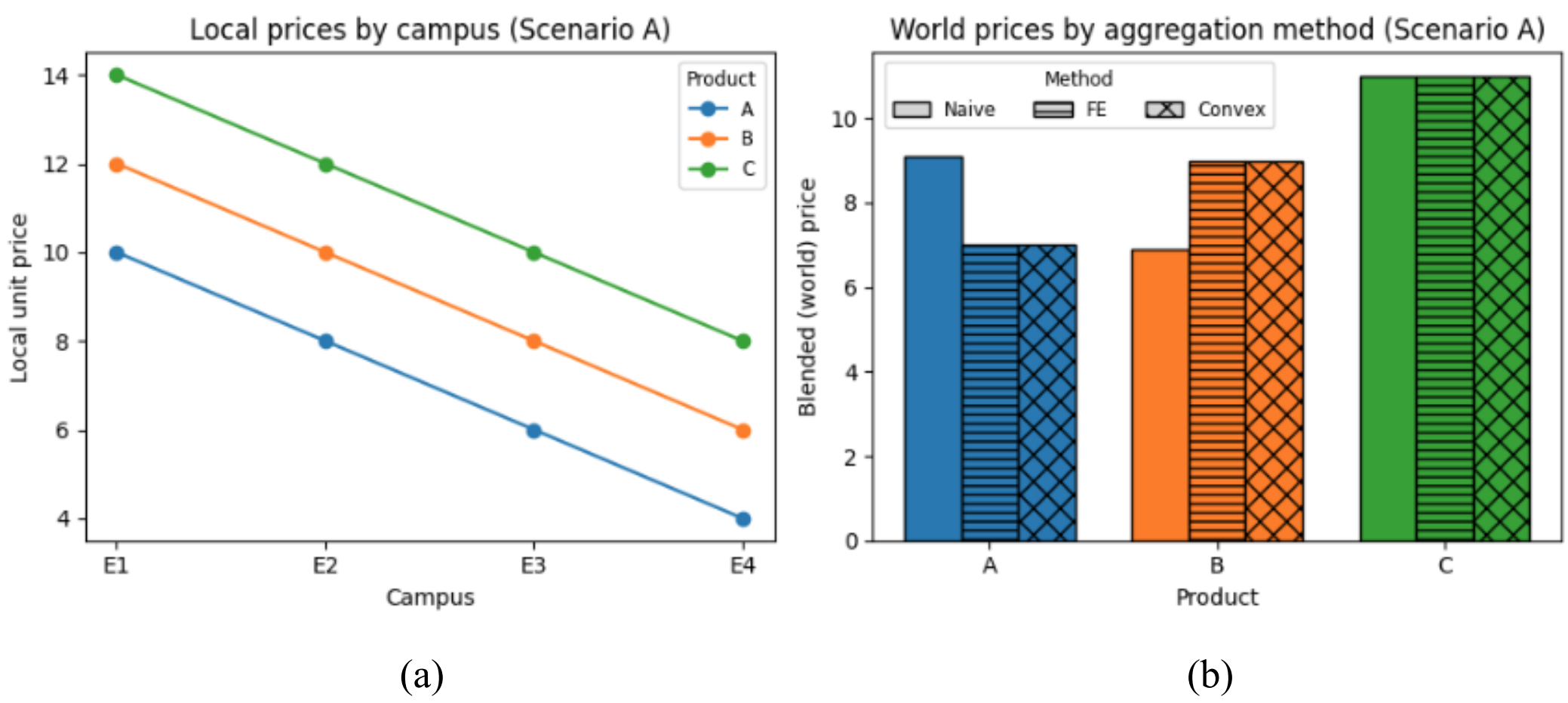

(a) (b)

*Notes:* Panel (a) plots local prices for three standardized trays A-C across four campuses E1-E4. The local ordering is strictly monotone and identical at all campuses, with A cheaper than B and B cheaper than C. Panel (b) shows the corresponding world prices from the naive, FE, and convex-weight aggregation operators. Naive blending, based on product-specific quantity weights, reverses the global ordering of A and B despite the common local ordering, illustrating a Simpson-type reversal. The FE and convex-weight operators restore a monotone world-price vector consistent with the local prices and maintain exact cost recovery.

*C.3 Scenario B: Realistic Mix Heterogeneity Beyond Dominance*

The second experiment enlarges the system to five products P1-P5 and eight campuses C1-C8. Panel (a) of Figure 3 again imposes a common local ordering $P1 < \cdots < P5$at all campuses, while allowing sizable cross-campus cost differences. As summarized in Table 2, naive blending now produces more subtle but still economically meaningful distortions. Because P2 is disproportionately deployed in cheaper campuses, its naive world price (11.10) falls below that of P1 (13.11), despite P1 being locally cheaper everywhere. More generally, products with a higher share of capacity in low-cost campuses are pulled down on the world-price scale, and those concentrated in high-cost campuses are pushed up, leading to an OVR of 0.30 for the naive operator. The FE and convex-weight indices again realign global prices with the underlying local hierarchy: both deliver strictly increasing world prices from P1 to P5, while being cost-preserving. Their OVR is zero and their CDRs are zero to numerical precision.

**Table C2.** Campus-level simulation, scenario B (J = 8 campuses, I = 5 products)

| Product | Naive world price | FE world price | Convex world price |
|---|---|---|---|
| P1 | 13.11 | 10.56 | 10.56 |
| P2 | 11.10 | 11.56 | 11.56 |
| P3 | 12.65 | 12.56 | 12.56 |
| P4 | 12.77 | 13.56 | 13.56 |
| P5 | 13.17 | 14.56 | 14.56 |
| OVR | 0.30 | 0.00 | 0.00 |
| CDR | 0.00 | 0.00 | 0.00 |

*Notes:* The table reports blended "world" prices for products P1-P5 across eight campuses under the naive, FE, and convex-weight operators. As in scenario A, campus-specific prices exhibit a common local ranking $P1 < \cdots < P5$, while quantity mixes differ by product. OVR measures the fraction of strictly dominant product pairs whose global ordering is violated by a given operator; CDR measures the proportional deviation between total realized cost and the cost implied by the corresponding world prices. All prices are rounded to two decimal places.

**Figure C2.** Campus-level local prices and blended world prices in scenario B

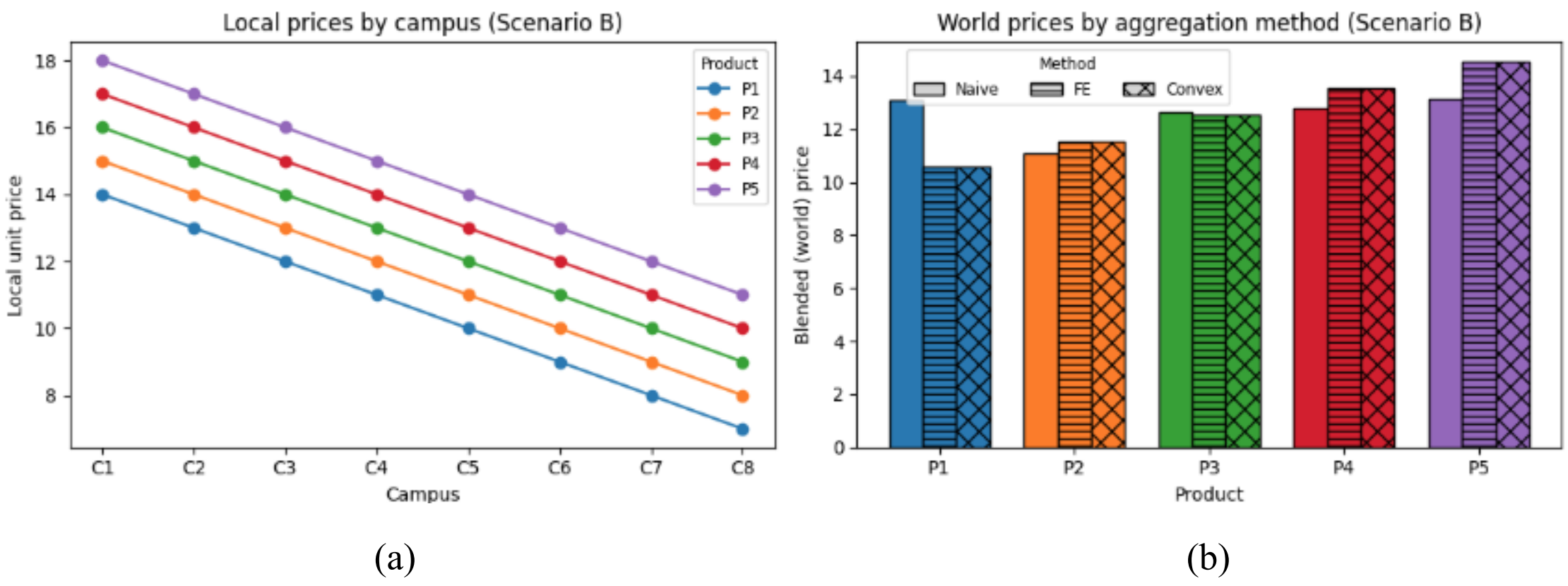

(a) (b)

*Notes:* Panel (a) displays local prices for products P1-P5 across eight campuses C1-C8 in scenario B. At every campus the local ordering is identical, with $P1 < \cdots < P5$, while price levels decline from C1 to C8. Panel (b) presents the blended world prices obtained from the naive, FE, and convex-weight operators. The naive index aggregates with product-specific quantity weights and therefore embeds heterogeneous deployment mixes into the global ranking. The FE and convex-weight indices impose additional structure on the weights and deliver world prices that are more closely aligned with the common local hierarchy while exactly preserving total cost.

All in all, these two campus-level simulations illustrate three key points. First, even when local prices are perfectly ordered across locations, heterogeneous quantity mixes are sufficient for naive blended prices to generate Simpson-type reversals and nontrivial OVR. Second, imposing either a two-way fixed-effects structure or a common convex weighting scheme is enough to eliminate these reversals in my examples, reducing OVR to zero while keeping CDR at essentially zero. Third, because all three operators are evaluated under identical local prices and quantity mixes, the improvement in OVR is achieved without any loss of cost recovery, highlighting that the proposed indices dominate naive blending on both ranking accuracy and cost-consistency criteria.

## References

[1] I. Fisher, The Making of Index Numbers: A Study of Their Varieties, Tests, and Reliability. Boston, MA, USA: Houghton Mifflin, 1922.

[2] W. E. Diewert, "Exact and superlative index numbers," J. Econom., vol. 4, no. 2, pp. 115-145, 1976. 10.1016/0304-4076(76)90009-9

[3] International Monetary Fund, International Labour Organization, OECD, UNECE, Eurostat, The World Bank, Consumer Price Index Manual: Concepts and Methods, 2020. Washington, DC, USA: IMF, 2020.

[4] B. M. Balk, Price and Quantity Index Numbers: Models for Measuring Aggregate Change and Difference. Cambridge, U.K.: Cambridge Univ. Press, 2008. 10.1017/CBO9780511720758

[5] E. H. Simpson, "The interpretation of interaction in contingency tables," J. R. Stat. Soc. Series B Stat. Methodological, vol. 13, no. 2, pp. 238-241, 1951. 10.1111/j.2517-6161.1951.tb00088.x

[6] C. R. Blyth, "On Simpson's paradox and the sure-thing principle," J. Am. Stat. Assoc., vol. 67, no. 338, pp. 364-366, 1972. 10.1080/01621459.1972.10482387

[7] J. Pearl, Causality: Models, Reasoning, and Inference, 2nd ed. Cambridge, U.K.: Cambridge Univ. Press, 2009. 10.1017/CBO9780511803161

[8] A. Shehabi, S. J. Smith, A. Hubbard, et al., 2024 United States Data Center Energy Usage Report, LBNL-2001637. Berkeley, CA, USA: Lawrence Berkeley National Laboratory, 2024.

[9] A. Shehabi, S. J. Smith, N. Horner, et al., United States Data Center Energy Usage Report. Berkeley, CA, USA: Lawrence Berkeley National Laboratory, 2016. https://eta-publications.lbl.gov/sites/default/files/lbnl-1005775_v2.pdf

[10] International Energy Agency, Energy and AI: Energy Demand from AI. Paris, France: IEA, 2025.

[11] V. Avelar, D. Azevedo, and A. French, PUE™: A Comprehensive Examination of the Metric. The Green Grid, 2012.

[12] Uptime Institute, 2024 Global Data Center Survey Report. Uptime Institute, 2024.

[13] McKinsey & Company, "How data centers and the energy sector can sate AI's hunger for power," 2024.

[14] S. Boyd and L. Vandenberghe, Convex Optimization. Cambridge, U.K.: Cambridge Univ. Press, 2004. 10.1017/CBO9780511804441

[15] M. Zaharia, M. Chowdhury, T. Das, et al., "Resilient distributed datasets: A fault-tolerant abstraction for in-memory cluster computing," in Proc. 9th USENIX Symp. Networked Systems Design and Implementation (NSDI), 2012, pp. 15-28.

[16] Q. He, "A Unified Metric Architecture for AI Infrastructure: A Cross-Layer Taxonomy Integrating Economics, Performance, and Efficiency," arXiv:2511.21772 [econ.GN], 2025. DOI: https://doi.org/10.48550/arXiv.2511.21772

[17] W. Sun, Q. Shen, Y. Gao, Q. Mao, T. Qi, and S. Xu, "Objective over Architecture: Fraud Detection Under Extreme Imbalance in Bank Account Opening," Computation, vol. 13, no. 12, p. 290, 2025, doi: 10.3390/computation13120290.

[18] Q. He and C. Qu, "Waste-to-Energy-Coupled AI Data Centers: Cooling Efficiency and Grid Resilience," arXiv:2512.24683 [eess.SY], 2025. DOI: https://doi.org/10.48550/arXiv.2512.24683

[19] Q. He and C. Qu, "Modular Landfill Remediation for AI Grid Resilience," arXiv:2512.19202 [eess.SY], 2025. DOI: https://doi.org/10.48550/arXiv.2512.19202

[20] Daniel J. Power, "Decision support systems: Concepts and resources for managers," Westport, CT: Quorum Books, 2002.

[21] D. Liu, Q. Shen, and J. Liu, “The Health-Wealth Gradient in Labor Markets: Integrating Health, Insurance, and Social Metrics to Predict Employment Density,” Computation, vol. 14, no. 1, p. 22, 2026, doi: 10.3390/computation14010022.

[22] Q. Shen and J. Zhang, “AI-Enhanced Disaster Risk Prediction with Explainable SHAP Analysis: A Multi-Class Classification Approach Using XGBoost,” Research Square, preprint, Dec. 2025, doi: 10.21203/rs.3.rs-8437180/v1.